\documentclass{pasj01}

\RequirePackage{mathpazo}
\RequirePackage[T1]{fontenc}

\newcounter{author}
\def\authorcount#1#2{\refstepcounter{author}\label{#1}
                     \altaffiltext{\ref{#1}}{#2}}

\begin{document}
\SetRunningHead{T. Kato}{Z Cam Stars with IW And-Type Phenomenon}

\Received{201X/XX/XX}
\Accepted{201X/XX/XX}

\title{Three Z Cam-Type Dwarf Novae
       Exhibiting IW And-Type Phenomenon}

\author{Taichi~\textsc{Kato}\altaffilmark{\ref{affil:Kyoto}*}
}

\authorcount{affil:Kyoto}{
     Department of Astronomy, Kyoto University, Kyoto 606-8502, Japan}
\email{$^*$tkato@kusastro.kyoto-u.ac.jp}



\maketitle

\begin{abstract}
I found that V507 Cyg, IM Eri and FY Vul
are Z Cam-type dwarf novae and
they showed sequences of standstill terminated by brightening,
in contrast to fading in ordinary Z Cam stars,
followed by damping oscillation.   These sequences
are characteristic to IW And-type objects (also known
as anomalous Z Cam stars).
New additions to the IW And-type objects suggests
that the IW And-type phenomenon is more prevalent among
Z Cam stars.  I suspect that the regularity of the pattern
of the IW And-type phenomenon suggests a previously
unknown type of limit-cycle oscillation, and I suggest that
the standstill in these objects is somehow maintained in
the inner part of the disk and the thermal instability
starting from the outer part of the disk terminates
the standstill to complete the cycle.
\end{abstract}

\section{Introduction}

   Dwarf novae are a class of cataclysmic variables,
which are close binary systems composed of
a white dwarf and a red-dwarf secondary
transferring matter via the Roche-lobe overflow.
In dwarf novae, thermal instability in the disk
is believed to cause repetitive outbursts.
[see e.g. \citet{osa96review};
for general information of cataclysmic variables
and dwarf novae, see e.g. \citet{war95book}].

   When the mass-transfer from the secondary is high,
the disk can become thermally stable (when the disk
is always thermally stable, such systems are called
novalike cataclysmic variables).
Z Cam stars are systems which transiently show
a thermally stable disk.
Standstills in Z Cam stars are intermediate states
between outburst maxima and minima, and Z Cam stars
spend states with ordinary dwarf nova outbursts and
standstills intermittently.  Z Cam stars are considered
to have relatively high mass-transfer rates
and the accretion disk is considered to be close to
the thermal stability.  It is generally considered
that a subtle variation in the mass-transfer rate
from the secondary causes transitions between
outbursting state and standstills \citep{mey83zcam}
[see also a review of dwarf nova outbursts
e.g. \citet{osa96review}].

   There has been at least one attempt to explain
standstills without introducing the mass-transfer
variation \citep{ros17zcam}.  This attempt has not
yet been successful to reproduce the actual Z Cam-type
behavior.

\begin{figure}
  \begin{center}
    \FigureFile(85mm,60mm){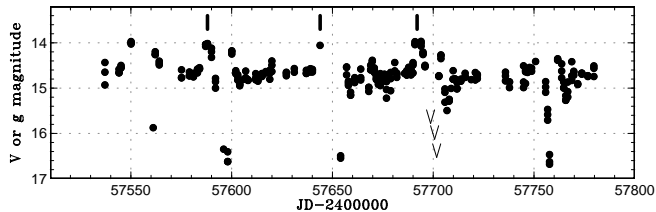}
  \end{center}
  \caption{Light curve of IW And using the ASAS-SN data.
  The ``V'' marks represent upper limits.
  The vertical ticks represents brightening terminating
  standstills.}
  \label{fig:iwandlc}
\end{figure}

   On the other hand, unusual Z Cam stars were
identified by the ``Z CamPaign'' led by \citet{sim11zcamcamp1}.
They were IW And and V513 Cas, which occasionally showed
a very regular sequence of standstill---brightening---damping oscillation
[figure \ref{fig:iwandlc} for IW And; 
figures 1 and 2 in \citet{sim11zcamcamp1}; also in
\citet{szk13iwandv513cas}].
The most interesting features are that (1) standstill are
terminated by brightening, not by fading as in ``textbook''
Z Cam stars, and (2) the sequence is often very regular with
almost a constant recurrence time ($\sim$40 d in IW And
in 2009--2010 and 40--50 d in V513 Cas in 2010).
Although these objects are sometimes referred to
as ``anomalous Z Cam stars'' (cf. \cite{ham14zcam}),
we propose a more specific name of ``IW And-type objects
(or phenomenon)'' after the brightest and easily pronounceable
prototype star.\footnote{The Z Cam-type nature of IW And
was first recognized by \citet{kat03iwand}, who incidentally
recorded an entrance to a standstill following outbursts.
A rise from the standstill in the final part of the observation
suggests that the anomalous ending of a standstill was
already present in 2002.}

   Another dwarf nova ST Cha was found to belong to
IW And-type objects \citep{sim14stchabpcra}.
Although the similarity with IW And-type objects was
not documented, the Kepler dwarf nova KIC 9406652
showed a similar pattern of outbursts (but with smaller
amplitudes) \citep{gie13j1922}.

   During the course of a systematic study of cataclysmic
variables using Gaia DR2 \citep{GaiaDR2}, I noticed
that variable stars V507 Cyg, IM Eri and FY Vul
belong to IW And-type stars.

\section{Observations}

   I used All-Sky Automated Survey for Supernovae (ASAS-SN)
Sky Patrol (\cite{ASASSN}; \cite{koc17ASASSNLC}).
These observations were made by using 14cm aperture
telephoto lenses and CCD cameras equipped with $V$ or $g$
filters [see \citet{koc17ASASSNLC} for more
technical details].  Observations were usually made
two or three times in one night and they were repeated
with intervals of 1--5 d (the intervals were sometimes
longer due to unfavorable conditions).
Since $V$ and $g$ magnitudes are almost the same in
dwarf novae, I used them without corrections to draw
light curves.

\section{Results}

\begin{figure}
  \begin{center}
    \FigureFile(85mm,110mm){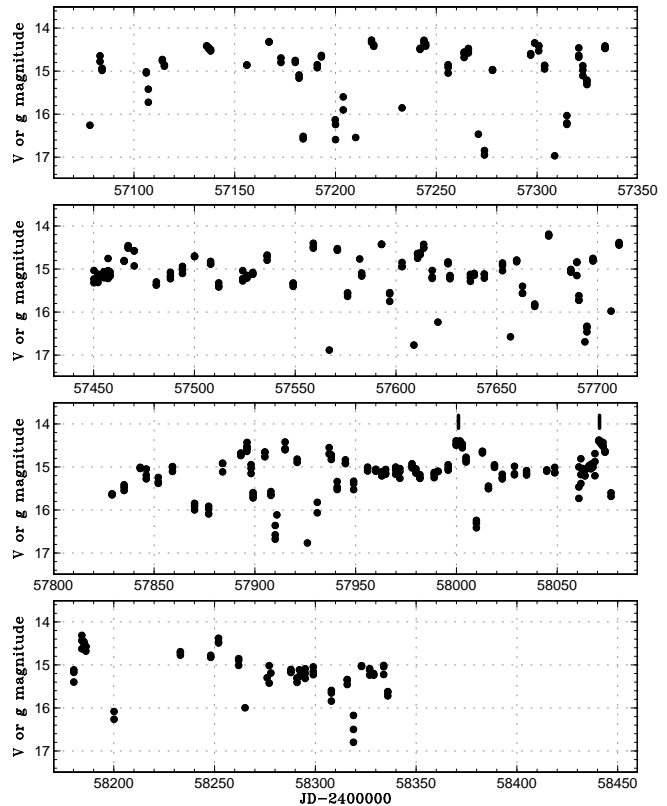}
  \end{center}
  \caption{Light curve of V507 Cyg using the ASAS-SN data.
  Two cycles of standstill---brightening---damping oscillation
  (IW And-type phenomenon)
  can be seen between JD 2457950 and JD 2457080.
  The vertical ticks represents brightening terminating
  standstills.}
  \label{fig:v507cyglc}
\end{figure}

\begin{figure}
  \begin{center}
    \FigureFile(85mm,80mm){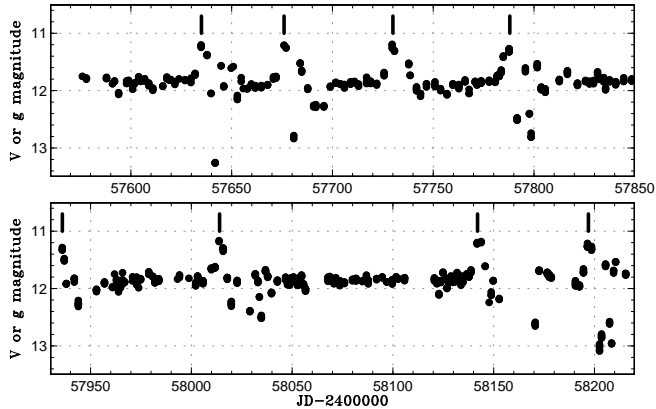}
  \end{center}
  \caption{Light curve of IM Eri using the ASAS-SN data.
  Cycles of standstill---brightening---damping oscillation
  (IW And-type phenomenon)
  can be seen throughout these observations.
  The vertical ticks represents brightening terminating
  standstills.  A dip after the third brightening
  (JD 2457730) was probably missed due to the gap in
  the observation.}
  \label{fig:imerilc}
\end{figure}

\begin{figure}
  \begin{center}
    \FigureFile(85mm,80mm){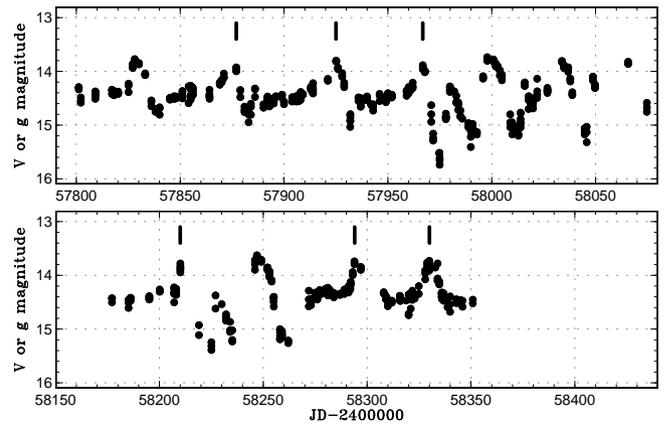}
  \end{center}
  \caption{Light curve of FY Vul using the ASAS-SN data.
  Somewhat less regular cycles of
  standstill---brightening---damping oscillation
  (IW And-type phenomenon) can be seen, particularly
  in the upper panel.
  The vertical ticks represents brightening terminating
  standstills.  The final two standstills were associated
  by brightening, but not followed by dips.}
  \label{fig:fyvullc}
\end{figure}

   The resultant light curve is shown in figure
\ref{fig:v507cyglc}.  Although the object spent much
of time showing typical Z Cam-type behavior with
occasional standstills (first and second panels),
the object showed the IW And-type phenomenon
(standstill---brightening---damping oscillation)
between JD 2457950 and JD 2457080 in the third panel.
IM Eri showed the same
phenomenon (figure \ref{fig:imerilc}).
The light curve of FY Vul is shown in figure
\ref{fig:fyvullc}.

\begin{figure}
  \begin{center}
    \FigureFile(85mm,60mm){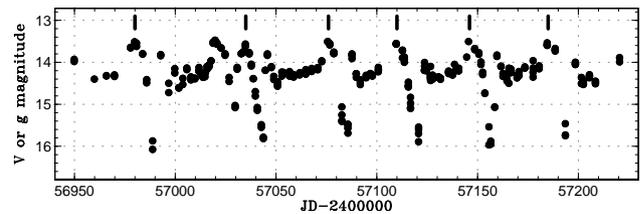}
  \end{center}
  \caption{Light curve of ST Cha using the ASAS-SN data.
  The IW And-type phenomenon was most prominent
  in this segment.
  The vertical ticks represents brightening terminating
  standstills.}
  \label{fig:stchalc}
\end{figure}

   For comparison, I provide the light curve of ST Cha
in the same scale in figure \ref{fig:stchalc}
since the regular nature of the IW And-type phenomenon
was less clear in \citet{sim14stchabpcra}.

\section{Discussion}

   With a new addition of members to the IW And-type
objects (and clear representation of the case of ST Cha
in this paper),
it has become more evident that the IW And-type
phenomenon is more prevalent among Z Cam stars
than previously thought.

   \citet{ham14zcam} suggested that termination of
standstills in the IW And-type phenomenon (standstills
terminate by brightening) can be explained by outbursts
of mass transfer with a duration of a few days.
The actual light curves of IW And-type objects,
however, are very regular (as seen in figure \ref{fig:stchalc}
and figures in \cite{sim11zcamcamp1}), and it would be
difficult to suppose such a regular occurrence of
``giant flares'' occurring near the secondary star surface
\citep{ham14zcam}.  The regularity, on the other hand,
would suggest a previously unknown limit-cycle oscillation
in the accretion disk [logically analogous to the one
proposed for the black-hole transient V404 Cyg in 
\citet{kim16v404cyg}].

\begin{figure}
  \begin{center}
    \FigureFile(85mm,60mm){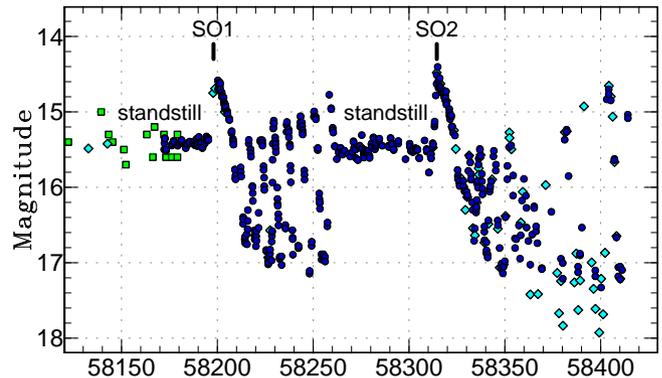}
  \end{center}
  \caption{Light curve of NY Ser in 2018 reproduced
  from figure 1 in \citet{kat18nyser}.  The abscissa
  is in BJD$-$2400000.
  This SU UMa-type dwarf nova showed standstills
  which were terminated by superoutbursts (SO1 and SO2).}
  \label{fig:nyserall}
\end{figure}

   Quite recently, we found that the SU UMa-type dwarf nova
NY Ser showed standstills and that superoutbursts started
from these standstills (figure \ref{fig:nyserall};
vsnet-alert 22026, 22313\footnote{
These vsnet-alert messages can be seen at
$<$http://ooruri.kusastro.kyoto-u.ac.jp/pipermail/vsnet-alert/$>$.
};
\cite{kat18nyser}).  These standstills and
superoutbursts were phenomenologically very similar
(standstills were terminated by brightening).
The development of superoutbursts during standstills
provides strong evidence that the disk radius reached
the 3:1 resonance during standstills in NY Ser.
This implies that the disk radius increased during
these standstills and eventually triggered the tidal
instability at the 3:1 resonance.  This phenomenon
in NY Ser suggests that the disk radius can increase
during standstills, possibly caused by re-distribution
of the surface density in the disk during standstills.
I consider that the similar situation could happen
in IW And-type objects: the standstill is initially
maintained in the inner part of the disk, and the disk
gradually expanded and thermal instability starting
from the outer part of the disk sweeps the entire disk
down to the quiescent state.

   The referee of this paper suggested a possibility
that a decreasing mass-transfer rate can lead to
a disk radius increase when the disk is in a high state
(as it is during a standstill).  I consider, however,
this interpretation less likely for several reasons:
(1) Although the disk can expand in response to
the decrease in the mass-transfer rate, the surface
density should decrease and the thermal instability
will become more difficult to start from the outer
disk.  (2) Several IW And-type stars show regular
cycles characterized by recurrence times of tens of
days.  This should require regular decrease in
the mass-transfer rate, which has not been observed
in other systems and would be difficult to produce.
(3) The brightness during standstills
increases in some systems (most clearly seen in FY Vul
and ST Cha), which contradicts the idea of decrease in
the mass-transfer rate.

   For these reasons, I prefer the original
working hypothesis that the thermal instability starting
from the outer part of the disk sweeps the entire disk
even with constant mass-transfer from the secondary
by a yet unidentified mechanism.  In FY Vul, brightening
from standstills did not result the subsequent quiescent state
(figure \ref{fig:fyvullc}).  I consider that the cooling
wave was not strong enough to sweeps the entire disk,
and that the system returned to the original standstill state.
It may be that the disk in standstills may be related to
the intermediate state in the `S-curve' of the thermal
equilibrium curve [\citet{mey87thermal};
figure 9 in \citet{osa14v1504cygv344lyrpaper3},
figure 6 in \citet{mey15suumareb}].
Further theoretical study is needed to explain how
the standstill can be maintained in
the inner part of the disk and why IW And-type phenomenon
is only seen in some Z Cam stars.

\section*{Acknowledgements}

The author is particularly grateful to the ASAS-SN team for
making their data available to the public.
A discussion with Y. Osaki about the IW And-type phenomenon
was very illuminating.
I acknowledge the referee Dr. Mattias Schreiber for
his constructive comments.
This research has made use of the SIMBAD database,
operated at CDS, Strasbourg, France.
This research has made use of the International Variable Star Index 
(VSX) database, operated at AAVSO, Cambridge, Massachusetts, USA.
This work has made use of data from the European Space Agency (ESA) mission
Gaia ($<$https://www.cosmos.esa.int/gaia$>$), processed by the Gaia
Data Processing and Analysis Consortium (DPAC,
$<$https://www.cosmos.esa.int/web/gaia/dpac/consortium$>$). Funding for the DPAC
has been provided by national institutions, in particular the institutions
participating in the Gaia Multilateral Agreement.
The author is grateful to H. Maehara for making Gaia DR2 data
available to the author enabling a systematic study of cataclysmic
variables.

\end{document}